# Successful Management of Cloud Based Global Software Development Projects: A Multivocal Study

**Muhammad Azeem Akbar[1][*], Sajjad Mehmood[2], Arif Ali Khan[3]**

[1]Lappeenranta-Lahti University of Technology, Department of Software Engineering, 53851, Lappeenranta, Finland.
[2]Information and Computer Science Department, King Fahd University of Petroleum and Minerals, Dhahran, Saudi Arabia.
[3]M3S Empirical Software Engineering Research Unit, University of Oulu, 90570 Oulu, Finland.

**Corresponding author:** Muhammad Azeem Akbar (Email: azeem.akbar@lut.fi)

**Abstract**
*Context:* Software industry is continuously exploring better ways to develop applications. A new phenomenon to achieve this is Cloud based Global Software Development (CGSD), which refers to the adoption of cloud computing services by organizations to support global software development projects. The CGSD approach affects the strategic and operational aspects of the way projects are managed. *Objective:* The objective of the study is to identify the success factors which contribute to management of CGSD projects. *Methods:* We carried out a Multivocal Literature Review (MLR) to identify the success factors from the state-of-the-art and the state-of-the-practice in project management of CGSD projects. We identified 32 success factors that contribute to the management of CGSD projects. *Results:* The findings of MLR indicate that time to market, continuous development, financial restructuring, scalability Moreover, the findings of the study show that there is a positive correlation between the success factors reported in both formal literature and industry based grey literature. *Conclusion:* The findings of this study can assist the practitioners to develop the strategies needed for effective project management of CGSD projects.

**Keywords:** Cloud based global software development, success factors, multivocal review, project management.

**1. Introduction**
A number of software development methods have been proposed to help organizations in application development ranging from traditional waterfall to agile methods[1]. The emergence of cloud sourcing facilitates organizations to have on-demand access to a shared and scalable range of IT resources such as storage and applications [2, 3]. In this paper, we adopt Schhneider and Sunyaey's definition of cloud scouring [3]: *"an organizations' decision to adopt and integrate cloud services from external providers into their IT landscape, that is, the customer organization's assessment of cloud computing offering from one or more providers in form of service model (IaaS, PaaS, SaaS) or deployment, model (e.g. public, private, community, hybrid)"*. Cloud sourcing changes how organizations manage their computing infrastructure by allowing them to transfer applications, services and data to cloud servers.

The success of cloud sourcing also captured the attention of the number of global software development organizations that strive to take advantage of scalability, on-demand services, and a large amount of virtual storage during the transformation of development activities to the distributed sites across the globe [4, 5]. This type of organizations leveraging cloud computing services to support the outsource development process is referred to as Cloud based Global Software Development (CGSD) [4]. With CGSD, organizations use on-demand access to a pool of scalable IT resources; and potentially increase product quality by having access to relatively low cost skilled human resources [6]. Moreover, CGSD has the potential to reduce the development time using follow-the-sun concepts [7] supported by different cloud service and deployment models.

Managing CGSD projects are challenging due to concerns related to security and privacy of data, 24/7 development model, and coordination complexities because of language, terminology, and cultural differences between geographically distributed teams [8-10]. CGSD also requires organizations to adjust their management processes [3] due to different service and deployment models. However, little empirical insight is available about the management decisions through which CGSD can be successful. A better understanding of the success factors associated with CGSD project management can be helpful to practitioners for carrying out project management activities in CGSD context.

The objective of this study is to identify the factors which impact successful management of CGSD projects. We also analyze the identified success factors from both client and vendor perspective. Moreover, the success factors are mapped into knowledge areas of the PMBOK. We believe that the identified success factors will assist the practitioners in developing the strategies for the successful managing the global software development activities using the cloud environment. We used the multivocal literature review (MLR) approach to collect evidence from both the state-of-the-art and practice in the management of CGSD projects. This study provides both scholars and practitioners with a knowledge base to identifying the success factors that contribute to the successful management of CGSD projects. In this paper, we address the following research questions:

RQ1: What are the success factors for managing CGSD projects as identified in the multivocal literature review?
RQ2: Are there any differences between the success factors identified in the formal and grey literature?
RQ3: What success factors are related to CGSD vendor or client organizations?
RQ4: What success factors are related to 10 knowledge areas of PMBOK?

In RQ1, we aim to identify the success factors that positively impact the cloud based global software development activities as reported by state-of-the-art and state-of-the-practices. In RQ2, we aim to study any potential differences or similarities between the success factors reported in both types of literature. In RQ3, we investigate the identified success factors from both client and vendor organizational perspective. Lastly, in RQ4, we map the identified CGSD success factors against 10 knowledge areas of PMBOK.

The remaining paper is structured as follows: related work and motivation of the study are presented in Section 2. Section 3 presented the research methodology. The results are discussed in Section 4. Section 5 presents the results summary, implications and future work and limitations of the study. Finally, the conclusion and future research directions are presented in Section 6.

## 2 Related Work and Study Motivation

Cloud service providers offer software services as "infrastructure as service," "platform as service," and "software as service" [11]. Moreover, cloud sourcing supports utility computing with different deployment options such as "private," "public," "hybrid," and "community" models [11-14]. Global software development is facing challenges with respect to business uncertainty, computing capacity, storage and security concerns [12-14]. Cloud sourcing is a promising solution to these challenges as it provides access to a shared pool of IT resources, applications, and services on a pay-per-use basis [15-17].

Several researchers reported how cloud sourcing can address some of the key challenges face by the global software industry. For example, Rousan [12] conducted a study to explore the critical challenges faced in cloud based GSD and presented a risk management model, which assists in analyzing and assessing the impact of risks. The model aims to analyze and prioritize cloud risks based on their impact on the GSD process. Similarly, Cocco et al. [13] presented a model for analyzing how GSD can be facilitated using a cloud environment. They developed a tool for small and medium software development organizations to determine the development time and cost while adopting cloud-based GSD paradigm. The tool provides a comparison between traditional and cloud-based GSD projects in terms of development time and cost. The study indicates that the cloud-based GSD is cost-effective and time-effective with respect to the traditional development paradigm. Furthermore, project managers can also use the tool to manage cloud-based GSD projects.

In another study, Yara et al. [14] highlighted that the cloud based GSD paradigm is significant for both client and vendor organizations as it helps organizations to overcome challenges with respect to computing capacity, storage and business uncertainty. They presented a generic cloud architecture to support GSD projects. Furthermore, they also highlighted that vendor lock-in, privacy, data access and regulatory compliance are key challenges faced by cloud-based GSD projects. Moreover, Haig-Smith and Tanner [18] used the domestication theory to understand how cloud sourcing was used in agile global software development. The study indicates that cloud sourcing can help reduce feedback latency in agile global software development projects. Moreover, they also found that cloud sourcing helps software developers in an agile global software project to focus on project delivery.

Oza et al. [19] conducted a qualitative study to identify the benefits and risks associated with cloud based distributed software development paradigm. The study mentioned that the cloud platform plays a significant role in managing the software development activities conducted in a distributed environment. Besides, they also reported that cloud platforms can be used to address the inherent challenges associated with global software development, such as temporal differences and task synchronization. Alajrami et al. [20] presented a software development as a service architecture to facilitate GSD projects in the cloud. Moreover, Alajrami et al. [21] integrated global distance metric into software development as a service architecture to provide task allocation decision support for GSD projects.

One key issue faced by global software development is that organizations try to incorporate cloud sourcing in their projects prior to understanding their management readiness. To date, no study has been conducted that presents evidence from both formal literature and grey literature to identify the success factors of managing CGSD projects. We believe there is a need for a study for both researchers and practitioners to understand success factors for managing CGSD projects from literature and industry point of view. This study presents the success factors which can assist organizations to manage CGSD projects.

## 3. Research methodologies

This study aims to identify the factors that contribute to project management success in CGSD projects. We used the Multivocal Literature Review (MLR) [22] methods, which is a type of systematic literature review that provide insight with respect to both the state-of-the-art and state-of-the-practice to answer the research questions [22-24]. As part of MLR, we reviewed the literature published in both formal (e.g., peer-reviewed journals, conferences etc.) and grey (industry standards, white papers, videos, blogs etc.) literature. Figure 1 presents an overview of our research methodology.

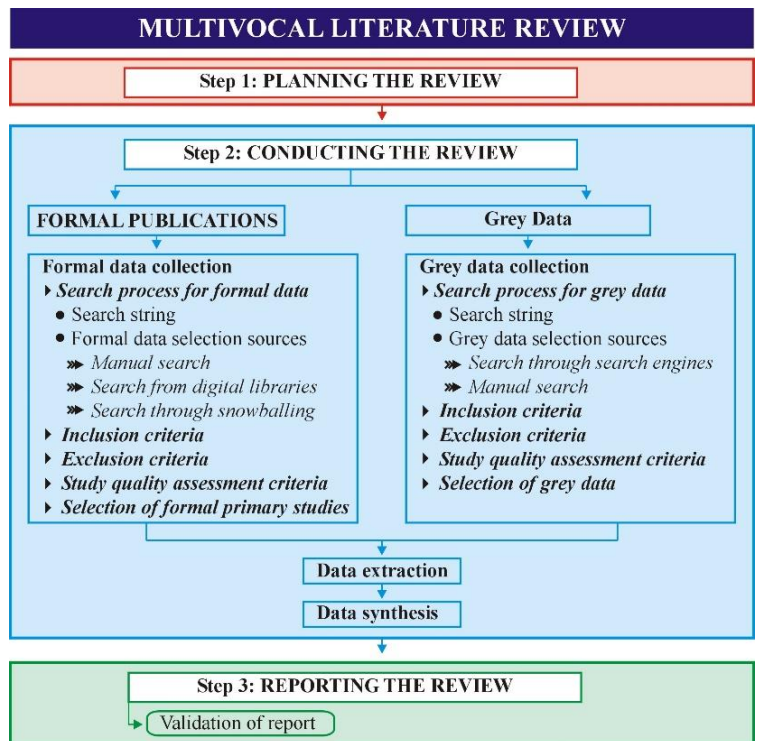

Figure 1: Proposed research methodology

### 3.1 Planning the review

As outlined in guidelines for MLR[22], a formal MLR protocol was developed before conducting the review process. All members of the research team participated in the MLR phases. Using the

MLR approach, data is collected from two sources i.e., formal published literature and grey literature. Both the data extraction processes are discussed in the following sections.

### 3.2 Data collection
#### 3.2.1 Formal data collection
To collect the peer-reviewed published primary studies, we followed the following steps:

*Search string*
We applied the guidelines provided by Zhang and Babar [25] to develop the search strings by concatenating the keyword and their alternatives. The keywords were selected by following the Quasi-Gold Standard (QGS) guidelines [26]. As part of the QGS guidelines, seven primary studies [FS1-FS7] (given in appendix-A) were selected to extract potential keywords to develop the search string. The keywords and their synonyms were concatenated using the Boolean "OR" and "AND" operators to structure the final search string as follows:

"(success factors OR factors OR aspects OR items OR elements OR drivers OR motivators OR variables) AND (Outsourcing OR global software development OR geographically distributed development OR offshore development OR multisite development OR collaborative software development) AND (IaaS OR PaaS OR SaaS OR XaaS OR infrastructure as a Service OR platform as a service OR Software as a Service OR IT service OR Application Service OR ASP) AND (cloud sourcing OR cloud computing OR cloud platform OR cloud provider OR cloud service OR cloud offering) AND (client software development organizations OR client software development companies OR  client analysis OR client perspective) AND (vendor software development organizations OR vendor software development companies OR vendor analysis OR vendor perspective) AND (client-vendor analysis OR client-vendor perspective)".

The following well known digital repositories are selected to execute the search strings and explore the results:

- "IEEE Xplore"
- "ACM Digital Library"
- "Springer Link"
- "Wiley Inter-Science"
- "Science Direct"
- "Google Scholar"
- "IET digital library"

Furthermore,  we also used forward and backward snowballing approaches to search the required data [27, 28]. The forward snowballing refers to the studies that cited the paper and the backward refer to the studies that are cited in the paper (reference list of the paper) [29, 30].  The snowballing is an effective way to collect the most potential literature related to the context of the study[29, 30].

*Inclusion criteria*
We developed the inclusion criteria based on the guidelines given in [9, 31, 32]. The key points of inclusion criteria include: "(i) study must be a conference paper, book chapter or journal article.

(ii) study should be about CGSD, (iii) study should contain the factors that could positively impact the management of CGSD projects, and (iv) results of the study should base on empirical evaluation".

*Exclusion criteria*

We applied the exclusion criteria developed to exclude the irrelevant studies based on the guideline given in [30, 32, 33]. The key points of exclusion criteria are: "(i) studies that were not relevant to the research objectives of this study (ii) studies that have not provided any details of the success factors that are significant for CGSD program (iii) studies were also excluded that were not written in English and (iv) the duplicate studies were also not considered".

*Study Quality Evaluation Criteria*

The quality of the selected studies are assessed using the criteria adopted from different other SLR studies in software engineering domains [9, 31, 34, 35]. The assessment criteria consist of the checklist that contains five quality assessment questions (Table 1). Each selected study was assessed against the questions of the assessment criteria using the given Likert scale (Table 1). The final score of each study is given in Appendix-A.

Table 1: quality assessment criteria formal primary studies for

| Questions of QA | Likert scale |
| --- | --- |
| "Does the research method used by the selected study address the RQs?" | "Yes=1, Partial=0.5, NO=0" |
| "Does the study, discuss any factor of CGSD?" | "Yes=1, Partial=0.5, NO=0" |
| "Does the study discuss the services of CGSD?" | "Yes=1, Partial=0.5, NO=0" |
| "Is the collected data related to the CGSD process?" | "Yes=1, Partial=0.5, NO=0" |
| "Are the findings of the study contribute to address the RQs?" | "Yes=1, Partial=0.5, NO=0" |

*Selection of formal primary studies*

Three different steps are followed to select the most relevant studies. The study selection steps are based on the suggestions provided by [30]. Initially, seven studies are manually selected by considering the suggestions of QGS [26]. The manually selected studies are directly related to the subject of the study. Moreover, we used the search strings discussed in section-3.2.1 and extracted 2273 studies from seven digital repositories using the inclusion and exclusion criteria. Furthermore, the tollgate approach [30] is used and in total, 132 primary studies are selected for the study, as shown in Figure 2. Every selected primary study (Appendix-A) is labeled as [FS] to indicate it as the MLR study.

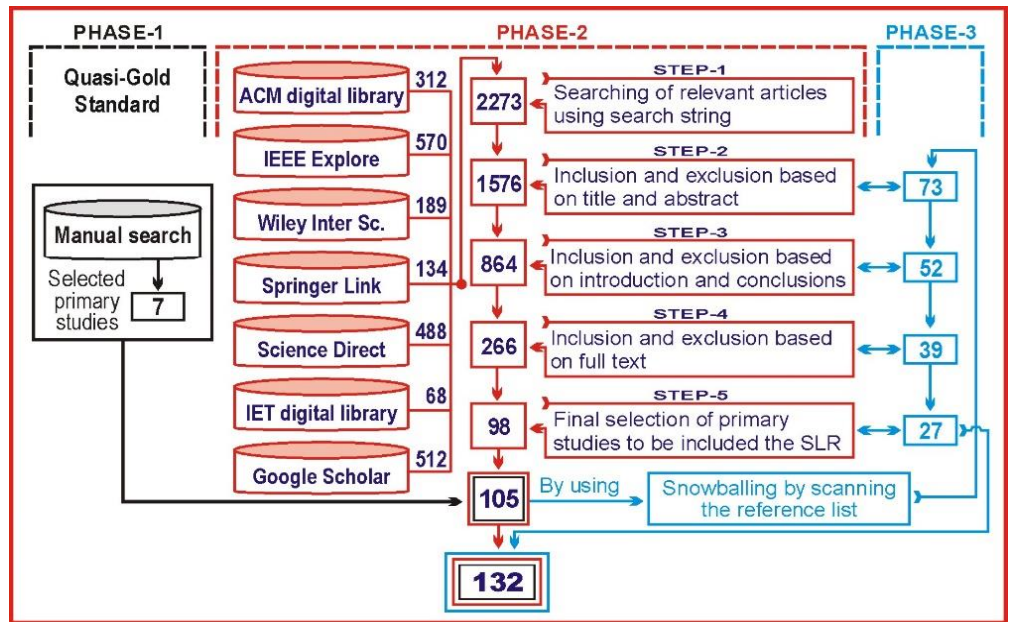
Figure 2: Final Selected Primary Studies

*3.2.2 Grey data collection*
Grey literature refers to the literature produced by government organizations, standards, business and industry which is not peer reviewed [36, 37]. We collected the grey (unpublished or non-peer reviewed) data based on the following steps:

*Search process for grey data*
The grey data are collected using the different search engines, websites and conducting manual search. The selected approaches are discussed as follow:

*Search through search engines*
To explore the grey literature, we used nine different digital sources. The selected sources include both the generic search engines and specialized libraries and websites. The search sources are selected by considering the suggestions provided by Garousi et al. [22] McGrath [38], Adams[39]. The following digital sources are carefully explored by executing the developed search string (section 3.2.1).

- "http://www.google.com"
- "https://www.bing.com"
- "http://www.opengrey.eu"
- "https://www.arxiv.org"
- "https://www.stackoverflow.com"
- "https://www.agilealliance.org"
- "https://www.istqb.org"
- "https://www.idc.com"

Moreover, the grey literature was also collected from the professional social media networks such as ResearchGate (https://www.researchgate.net) and LinkedIn (https://www.linkedin.com). We also approached different organizations through email and requested the data relevant to the research questions of this study. The personal and professional Email IDs were collected from the organization websites and through our supervisor's contacts in the software industry. We joined different available social media professional groups and send requests to provide unpublished data e.g., research notes, research registers, case studies, videos, ppt. presentations, etc. The respondents were assured that the collected data will only be used for research purposes and will keep confidential. The demographic details of the respondents are provided in Appendix-C.

*Inclusion criteria for grey data*
The following inclusion criteria for grey literature were developed based on the guidelines presented in [22].
(i) The study is relevant to the objectives of the study.
(ii) The study provides details about success factors of managing cloud-based software development outsourcing.
(iii) The study provides contextual information about the subject under study.
(iv) The study is useful and evident for both industrial and academic researchers.

*Exclusion criteria for grey data*
The following exclusion criteria for grey literature were developed based on the guidelines presentation in [22].
(i) The study does not provide details on the management of CGSD.
(ii) The study does not provide empirical evidence.
(iii) The study was not in the English language.

*Quality assessment criteria for grey literature*
The quality assessment process is performed to check the degree of credibility of the selected grey data. It is important to assess the quality of the data to measure its significance related to the study objectives. We develop a QA checklist for evaluating the significance of the grey data based on the suggestions presented in [22]. The checklist along the QA questions is provided in Table 2. The selected grey literature data are assessed based on the given checklist questions and the detail results are provided in Appendix-B.

Table 2: Quality assessment criteria for grey literature adopted from [22]

| Questions of QA | Likert scale |
| --- | --- |
| "Is the publishing organization reputable? E.g., the Software Engineering Institute (SEI)" | "Yes=1, Partial=0.5, NO=0" |
| "Is an individual author associated with a reputable organization?" | "Yes=1, Partial=0.5, NO=0" |
| "Has the author published other work in the field?" | "Yes=1, Partial=0.5, NO=0" |
| "Does the author have expertise in the area? (e.g., job title principal software engineer)" | "Yes=1, Partial=0.5, NO=0" |
| "Does the source have a clearly stated aim?" | "Yes=1, Partial=0.5, NO=0" |
| "Does the source have a stated methodology?" | "Yes=1, Partial=0.5, NO=0" |
| "Is the source supported by authoritative, contemporary references?" | "Yes=1, Partial=0.5, NO=0" |
| "Are any limits clearly stated?" | "Yes=1, Partial=0.5, NO=0" |
| "Does the work cover a specific question?" | "Yes=1, Partial=0.5, NO=0" |
| "Does the work refer to a particular population or case?" | "Yes=1, Partial=0.5, NO=0" |
| "Does the work seem to be balanced in the presentation?" | "Yes=1, Partial=0.5, NO=0" |
| "Is the statement in the sources as objective as possible? Or is the statement a subjective opinion?" | "Yes=1, Partial=0.5, NO=0" |

| "Is there vested interest? E.g., a tool comparison by authors that are working for a particular tool vendor." | "Yes=1, Partial=0.5, NO=0" |
|---|---|
| "Are the conclusions supported by the data?" | "Yes=1, Partial=0.5, NO=0" |
| "Does the item have a clearly stated date?" | "Yes=1, Partial=0.5, NO=0" |
| "Have key-related GL or formal sources been linked to/ discussed?" | "Yes=1, Partial=0.5, NO=0" |

**Grey data collection process**

The search strings were executed in the selected search engines to extract the most relevant and appropriate grey data. A total of 312 websites, blogs, and webpages were summarized after applying the data inclusion (section 3.2.2) and exclusion (section 3.2.2) criteria. We further filter the data sources using the tollgate approach developed by Afzal et al. [30]. The step-by-step process of the tollgate approach is given in Figure 3. After applying the five phases of data refinement process (Figure 3), finally, we collect the 37 data sources. Besides, we apply manual search (as discussed section 3.2.2), and a total of 27 responses are received from the practitioners. We apply the similar steps of Figure 3 to refine the data collected from the practitioner and finally, 15 responses were considered to include in the data extraction process. Thought, through the grey data collection process, a total of (37+15=52) data sources were included in MLR. We provide the links of the webpages (Appendix-B) and the information of data collected through personal contact with practitioners in Appendix-C. The personal contact should remain confidential according to the recommendation of respondents. The quality assessment process (i.e., discuss in section 3.2.2) is performed for all the data sets and the results are provided in Appendix-B and Appendix-C respectively. Every selected data source is labeled with "GL" to indicate its use in MLR.

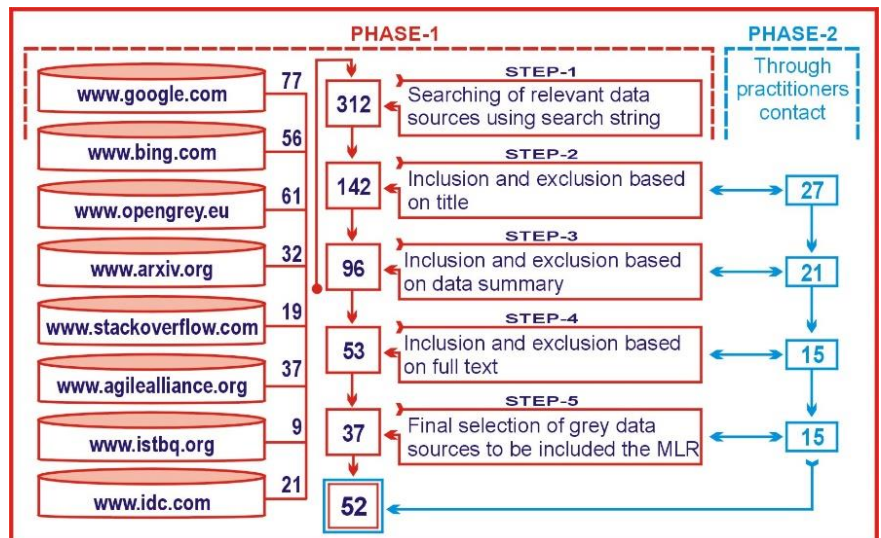

Figure 3: Final selected grey data by employing the tollgate approach

### 3.2.3 Data extraction

We used the coding approach [40] to analyze the grey data and identify the success factors. The identified ideas, concepts, contribution or findings from the selected data were labeled, grouped and classified as the success factors with their respective frequency. The coding approach assists in analyzing the analytical data in which the success factors were identified with deep and effective analysis of qualitative data sets.

Once the data extraction process is finalized, we conducted an inter-rater reliability test to examine the interpersonal bias. To do this, we request the three independent experts, and they selected 15 data sources (10 published primary studies and five sources of grey literature). They performed all the phases of data collection and data extraction. We determine the "Kendall's nonparametric coefficient of concordance (W)" to measure the inter-rater agreement [30] between the data extraction team and independent experts. The result, W = 0.84 (p = 0. 003), demonstrate agreement among the data extraction team and the independent experts, as shown in Table 3.

Table 3: "Kendall's coefficient of concordance test"

| "Data Set" | "Kendall Chi-Squared" | "df" | "Subjects" | "Raters" | "$p$ value" | "W" |
|---|---|---|---|---|---|---|
| CSOD | 35.434 | 14 | 15 | 3 | 0.001267 | 0.8436765 |

### 3.2.4 Data synthesis
Finally, we identify success factors based on the data extracted from formal studies and grey literature.
The research questions were assessed against the extracted data from both data sources. The identified success factors are discussed in section 4.

### 3.3 Reporting the review
### 3.3.1 Quality assessment of primary selected studies.
The quality of the selected data sources (primary studies and grey literature) was assessed to assure the degree of effectiveness of each data source. The quality assessment checklist discussed in Table 1 and 2 were applied to selected data sets. The QA results show that 79% of the primary studies scored greater than 70%, and 93% of grey literature scored≥ 70%. The detailed QA results are provided in Appendix-A (formal primary studies), Appendix-B and Appendix C (grey literature).

### 3.3.2 Review results
The identified success factors of the CGSD process and the additional analysis is discussed in section 4.

## 4. Results and discussions
This section contains the findings of the MLR study.

### 4.1 Identified success factor of CGSD
Using the MLR approach, a total of 32 factors that have a positive impact on cloud-based outsourcing are identified from the selected data sources (formal and grey literature) as presented in Table 4.

Table 4: List of investigated success factor

| Factors ID | Factors | F (N=184) | Percentage |
|---|---|---|---|
| SF1 | Integration with organizational IT infrastructure | 56 | 30 |
| SF2 | Continuous development | 102 | 55 |
| SF3 | Time to market | 117 | 64 |
| SF4 | Real-time tracking and traceability | 81 | 44 |
| SF5 | Financial restructuring | 93 | 51 |
| SF6 | Cost assessment and budget allocation | 62 | 34 |
| SF7 | Requirements change management | 41 | 22 |

| SF8  | Scalability | 92 | 50 |
| SF9  | Work dynamics | 74 | 40 |
| SF10 | Incremental cycles | 52 | 28 |
| SF11 | Consistent quality of services | 61 | 33 |
| SF12 | Trust building | 96 | 52 |
| SF13 | Better employee access to information and applications | 77 | 42 |
| SF14 | Apply the Right 3P (People, Processes and Partners) | 82 | 45 |
| SF15 | Formalize relationship between overseas teams | 111 | 60 |
| SF16 | Knowledge sharing | 54 | 29 |
| SF17 | Uniform communication infrastructure | 69 | 38 |
| SF18 | Information safety and security | 114 | 62 |
| SF19 | Master storage disk of primary data | 58 | 32 |
| SF20 | Legislation and regulation with cloud service provider | 38 | 21 |
| SF21 | Choose the right cloud service provider | 44 | 24 |
| SF22 | Designing an enterprise IT architecture for cloud services | 61 | 33 |
| SF23 | Greater resource agility | 40 | 22 |
| SF24 | Reliable support structure | 100 | 54 |
| SF25 | Technology Catalyst | 47 | 26 |
| SF26 | Business transition | 49 | 27 |
| SF27 | Business innovation | 58 | 32 |
| SF28 | Flexible governance | 49 | 27 |
| SF29 | Hardware and infrastructure independence | 56 | 30 |
| SF30 | Customer awareness | 73 | 40 |
| SF31 | Client firm IT capabilities | 37 | 20 |
| SF32 | Client and vendor interaction | 76 | 41 |

*SF3 (Time to market, 64%)* has highest citation frequency. There is a need for "agility" in terms of being able to understand and adapt the trend information and new development techniques [FS17]. For example, an organization can use a business intelligence suite to get real-time information on product sales, which allows management to adjust their strategy. In the CGSD environment, the development teams are in a geographically distributed environment that assists in predicting the current market trend. Moreover, Weinhardt et al. [FS27] stated that CGSD, the distributed teams work parallels to develop the system components that minimize the time need to deliver the product in the market.

*SF18 ("Information safety and security", 62%)* was cited as the second highest frequency success factor for the management of CGSD. Yanosky et al. [FS30] emphasize significantly considering the parameters of data privacy in the contract signed with the cloud provider. They further suggested four key points manage security risks, i.e., password protection, remote access, encrypted data, network security, and backup data. Different other researchers have also indicated importance of information security in the CGSD environment [SF19, 43, 85]. Moreover, Jain [FS47] recommended managing the backup files of the data that used across the cloud platform.

*SF15 ("Formalize relationship between overseas teams", 60%)* was specified as the third most cited success factor. In the CGSD process, the development phases are carried across distributed locations [FS43]. However, the software development activities demand rich communication and coordination channels between CGSD the teams which are significant to share frequent and instant information. Yigitbasioglu et al. [FS67] highlighted that poor communication causes the misinterpretation of software requirements and expectations that could be a potential barrier while managing the CGSD activities. Corney et al. [FS56] and Gonzalez et al. [128] also mentioned the importance of a strong and trustworthy relationship between the CGSD teams.

*SF2 (Continuous development, 55%)* was indicated as a key success factor for the CGSD environment. Garrison et al. [FS40] shown that it is critical to manage the continuous and round the clock development of the software activities. Rouse and Corbitt et al. [FS97] indicated that different time zones in the distributed environment has both negative and positive impacts at the same time. They further underlined that the time differences in different zones provide a positive edge to continuously and parallels perform the software development activities.

*SF24 (Reliable support structure, 54%)* was also cited as a success factor for the successful execution of CGSD. To manage cloud services, an effective help desk should be established to help practitioners [SF39]. Gens [FS58] highlighted that for the maximum utilization of cloud services, the outsourced organization provides a skilled help desk to the practitioners. Cardoso [FS72] suggested both automated (messages on the systems tabs) and manual (support team) support desk. Bennani [FS61] stated that the demands and needs of the practitioners should address in-time and efficiently, which is important for the success of CGSD projects.

*SF5 (Financial restructuring, 51%)* Leimeister et al. [FS16] stated that the main aim of software outsourcing is to develop a product at a low cost and with high quality. Oliveira et al. [FS36] highlighted that in the CGSD environment, the cost factor should manage concerning the nature of the development life cycle. Similarly, Siepmann [FS69] highlighted that the CGSD paradigm provides technological independence, which attributed that the CGSD need to spend low cost on technological intrastate but pay more for controlling the outsourcing activities. They further stated that the CGSD organization effectively manages the budget to handle the other concerns such as hidden costs, data backup, frequent communication etc.

*SF12 (Trust building, 51%)* Trust of one of the key factors that are critically significant for managing software development activities using a cloud platform. Tsai et al. [FS64] and Yin et al. [FS60] suggested that frequent communication and coordination opportunities should be arranged between the overseas practitioners to develop trust and work relationships. Ojala and Tyrvainen [FS49] further specified that due to the lack of trust, the practitioners hesitate to share the project related information and knowledge that could significantly impact the three key parameters of the project: schedule, quality, and budget.

*SF8 (Scalability, 50%)* Scalability is a key feature of cloud infrastructure [FS120]. Nuseibeh [FS74] highlighted that the cloud platform provides an opportunity to upscale or downscale the cloud resources. Cloud providers allow the client organizations to scale the resources as per project needs and requirements. It will further allow CGSD firms to support business growth without expensive changes to their technological infrastructure [FS21, FS47, SF77].

### 4.2 Client and vendor analysis
A total of 82 data sources (published primary studies and grey literature) were considered in the domain of client and 102 for vendor organizations. The investigated success factors were categorized in client and vendor CGSD firms (Table 5). The Chi-square test was used to examine the significant differences between the investigated success factors from both client and vendor organizations' perspective. Similar analysis has been applied in similar studies (e.g. [9, 31, 33, 34, 41-45]).

- *Null hypothesis (H0):* There is no significant difference between the client and vendor CGSD organizations with respect to investigated success factors.
- *Alternate hypothesis (H1)*: There are significant differences between the client and vendor CGSD organizations with respect to investigated success factors.

Null hypothesis (H0) is accepted if the significance value of 'P' is >0.05 for any success factor; else, the alternative hypothesis H1.

Table 5: Client vendor classification

| S.NO | Success Factors | Client (N=82) | | Vendor (N=102) | | "Chi-square Test (Linear-by-Linear Association) α = 0.05" | | |
|---|---|---|---|---|---|---|---|---|
| | | F | % | F | % | $X^2$ | df | P |
| SF1 | Integration with organizational IT infrastructure | 21 | 26 | 35 | 34 | 1.442 | 1 | 0.230 |
| **SF2** | **Continues development** | **67** | **82** | **35** | **34** | **7.397** | **1** | **0.008** |
| SF3 | Time to market | 66 | 80 | 51 | 50 | 1.442 | 1 | 0.230 |
| SF4 | Real-time tracking and traceability | 41 | 50 | 40 | 39 | 0.037 | 1 | 0.848 |
| SF5 | Financial restructuring | 43 | 52 | 50 | 49 | 0.222 | 1 | 0.637 |
| SF6 | Cost assessment and budget allocation | 29 | 35 | 33 | 32 | 1.165 | 1 | 0.280 |
| SF7 | Requirements change management | 14 | 17 | 27 | 26 | 0.806 | 1 | 0.369 |
| SF8 | Scalability | 31 | 38 | 61 | 60 | 2.052 | 1 | 0.152 |
| SF9 | Work dynamics | 36 | 44 | 38 | 37 | 0.007 | 1 | 0.931 |
| SF10 | Incremental cycles | 25 | 30 | 27 | 26 | 0.123 | 1 | 0.726 |
| SF11 | Consistent quality of services | 28 | 34 | 33 | 32 | 0.066 | 1 | 0.797 |
| SF12 | Trust building | 33 | 40 | 63 | 62 | 2.576 | 1 | 0.108 |
| SF13 | Better employee access to information and applications | 22 | 27 | 55 | 54 | 0.022 | 1 | 0.881 |
| SF14 | Apply the right 3P (People, Processes and Partners) | 38 | 46 | 44 | 43 | 0.356 | 1 | 0.551 |
| SF15 | Formalize relationship between overseas teams | 59 | 72 | 52 | 51 | 1.927 | 1 | 0.165 |
| SF16 | Knowledge sharing | 22 | 27 | 32 | 31 | 0.006 | 1 | 0.938 |
| **SF17** | **Uniform communication infrastructure** | **17** | **21** | **52** | **51** | **5.713** | **1** | **0.041** |
| SF18 | Information safety and security | 57 | 70 | 57 | 56 | 0.157 | 1 | 0.692 |
| SF19 | Master storage disk of primary data | 25 | 30 | 33 | 32 | 0.229 | 1 | 0.633 |
| SF20 | Legislation and regulation with cloud service provider | 22 | 27 | 16 | 16 | 2.691 | 1 | 0.101 |
| SF21 | Choose the right cloud service provider | 23 | 28 | 21 | 21 | 0.160 | 1 | 0.689 |
| SF22 | Designing an enterprise IT architecture for cloud services | 21 | 26 | 40 | 39 | 0.270 | 1 | 0.603 |
| SF23 | Greater resource agility | 16 | 20 | 24 | 24 | 0.265 | 1 | 0.607 |
| **SF24** | **Reliable support structure** | **62** | **76** | **38** | **37** | **4.049** | **1** | **0.034** |
| SF25 | Technology catalyst | 17 | 21 | 30 | 29 | 1.178 | 1 | 0.278 |
| SF26 | Business transition | 26 | 32 | 23 | 23 | 0.976 | 1 | 0.323 |
| SF27 | Business innovation | 31 | 38 | 27 | 26 | 0.007 | 1 | 0.931 |
| SF28 | Flexible governance | 29 | 35 | 20 | 20 | 0.123 | 1 | 0.726 |
| SF29 | Hardware and infrastructure independence | 25 | 30 | 31 | 30 | 0.066 | 1 | 0.797 |
| SF30 | Customer awareness | 41 | 50 | 32 | 31 | 2.576 | 1 | 0.108 |
| SF31 | Client firm IT capabilities | 14 | 17 | 23 | 23 | 1.163 | 1 | 0.235 |
| SF32 | Client and vendor interaction | 31 | 38 | 45 | 44 | 2.576 | 1 | 0.108 |

The results of the chi-square test presented in Table 5 shows that there are significant differences for only three factors: *SF2 (continues development (around the clock), p=0.008)*, *SF17 (uniform communication infrastructure, 0.041)* and *SF24 (reliable support structure, 0.034)* and the Null hypothesis is only accepted for the given three factors. We noted that *SF2 (continuous development)* is highly considered in the client organization category. Tripathi and Parihar [FS66] mention that the client organizations outsourced their development activities across the world to minimize development time; therefore, they significantly considered the SF2. Hudaib et al. [FS31] and Bennan et al. [FS61] also highlighted that for improving the production efficiency, the client organizations should outsource their businesses to maximize the working hours. *SF17 (uniform communication infrastructure)* has the highest frequency of occurrence in vendor organizations. Yang [FS71] underlined that in the CGSD environment, the development activities are conducted in vendor organizations, and the practitioners need to frequently communicate and coordinate. Moreover, Dutta et al. [FS80] emphasized the provision of proper technological and social platforms to make the communication and coordination process more effective.

Moreover, SF5 *(financial restructuring, 52%, and 49%)*, SF6 *(cost assessment and budget allocation, 35%, and 32%)*, SF11 *(consistent quality of services, 34%, and 32%)*, SF14 *(apply the right 3P (people, processes and partners), 46% and 43%)*, SF19 *(master storage disk of primary data, 30% and 32%)*, SF29 *(hardware and infrastructure independence, 30% and 30%)*, are declared as the most common and equally important success factors in both client and vendor organization. The CGSD organization should strongly focus on the success factors which are highly cited in their respective organization type.

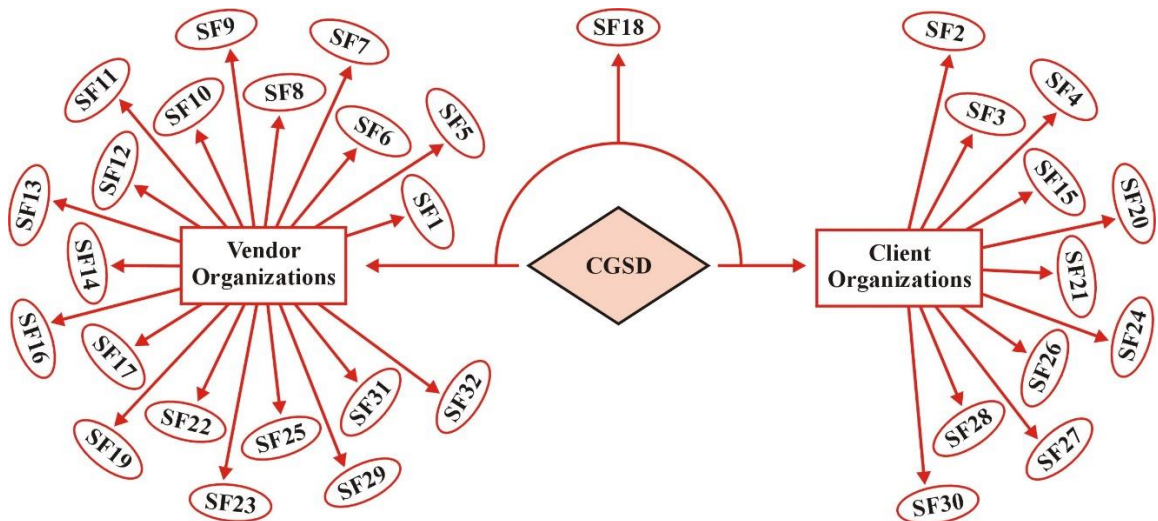

Figure 3: Mapping of success factors in Client-Vendor firms

We further categorized the CGSD success factors in client and vendor firms considering the framework of Ramasubu [46]. Ramasubu classify the GSD process improvement barriers in different domains. This conceptual mapping is also used by several other studies[31, 35]. Thus, we considered the same framework and map each success factor in both types of organizations, using their frequency of occurrence (Table 5) and the mapping results are given in Figure 3. For instance, SF1 (Integration with organizational IT infrastructure) is considered by 26% and 34% by client and vendor organization, respectively. As SF1 has higher frequency of occurrence in vendor firms' category, thus it is mapped with vendor organizations. Using the same concept, all the 32 identified success factors were mapped with client-vendor organizations and the mapping results are given in Figure 3.

According to the mapping results 20 success factors are aligned with vendor organizations and 11 are with client organizations. Interestingly, SF18 (Information safety and security) has equal frequency of occurrence in both types of organizations, this renders the SF18 is equally important for both types of organizations.

### 4.3 Comparison of Formal and Grey literature

We comparatively analyzed the two data sets (formal and grey literature) to evaluate the significant differences (Table 6). We calculate the rank of each factor based on the two data sets by taking the average value. The rank values of each factor were used to calculate the correlation between the two data sets. In this study, we used the Spearman rank-order correlation analysis to investigate the similarities and differences in both data sets [33].

Table 6: Ranks obtain from both data sets

| S.NO | Success Factors | Formal Literature | | | Grey Literature | | |
|---|---|---|---|---|---|---|---|
| | | F(n=132) | % | Rank | F (n=52) | % | Rank |
| SF1 | Integration with organizational IT infrastructure | 27 | 26 | 20 | 29 | 42 | 12 |
| SF2 | Continues development | 71 | 54 | 4 | 31 | 58 | 6 |
| SF3 | Time to market | 76 | 58 | 2 | 41 | 79 | 2 |
| SF4 | Real-time tracking and traceability | 55 | 42 | 10 | 26 | 50 | 9 |
| SF5 | Financial restructuring | 57 | 43 | 9 | 35 | 67 | 3 |
| SF6 | Cost assessment and budget allocation | 41 | 31 | 16 | 21 | 40 | 13 |
| SF7 | Requirements change management | 27 | 20 | 24 | 14 | 27 | 19 |
| SF8 | Scalability | 51 | 45 | 8 | 41 | 62 | 4 |
| SF9 | Work dynamics | 43 | 36 | 12 | 31 | 52 | 8 |
| SF10 | Incremental cycles" | 31 | 25 | 21 | 21 | 37 | 15 |
| SF11 | Consistent quality of services" | 49 | 37 | 11 | 12 | 23 | 21 |
| SF12 | Trust building | 57 | 51 | 6 | 42 | 56 | 7 |
| SF13 | Better Employee Access to Information And Applications | 46 | 35 | 13 | 31 | 60 | 5 |
| SF14 | Apply the Right 3P (People, Processes and Partners) | 62 | 47 | 7 | 20 | 38 | 14 |
| SF15 | Formalize relationship between overseas teams | 70 | 53 | 5 | 41 | 79 | 2 |
| SF16 | Knowledge sharing | 38 | 29 | 17 | 16 | 31 | 17 |
| SF17 | Uniform communication infrastructure | 46 | 35 | 13 | 23 | 44 | 11 |
| SF18 | Information safety and security | 87 | 66 | 1 | 27 | 52 | 8 |
| SF19 | Master storage disk of primary data | 36 | 27 | 19 | 22 | 42 | 12 |
| SF20 | Legislation and regulation with cloud service provider | 21 | 20 | 24 | 17 | 21 | 22 |
| SF21 | Choose the Right Cloud Service Provider | 33 | 25 | 21 | 11 | 21 | 22 |
| SF22 | Designing an enterprise IT architecture for cloud services | 42 | 32 | 15 | 19 | 37 | 15 |
| SF23 | Greater Resource Agility | 27 | 20 | 24 | 13 | 25 | 20 |
| SF24 | Reliable support structure | 79 | 55 | 3 | 21 | 48 | 10 |
| SF25 | Technology Catalyst | 32 | 24 | 22 | 15 | 29 | 18 |
| SF26 | Business transition | 37 | 28 | 18 | 12 | 23 | 21 |
| SF27 | Business innovation | 49 | 32 | 15 | 9 | 31 | 17 |
| SF28 | Flexible governance | 31 | 23 | 23 | 18 | 35 | 16 |
| SF29 | Hardware and Infrastructure Independence | 45 | 34 | 14 | 11 | 21 | 22 |
| SF30 | Customer awareness | 31 | 23 | 23 | 42 | 81 | 1 |
| SF31 | Client firm IT capabilities | 19 | 14 | 25 | 18 | 35 | 16 |
| SF32 | Client and vendor interaction | 49 | 37 | 11 | 27 | 52 | 8 |

Results of "Spearman's rank-order correlation" ($r_S$ (32) =0.612) shows the positive correlation between both data sets. Table 7 and Figure 4 show the results and scatter plot respectively.

Table 7: Correlation of both data sets

| | | | Formal_Literature | Grey_Literature |
|---|---|---|---|---|
| "Spearman's rho" | Formal_Literature | "Correlation Coefficient" | 1.000 | 0.612** |
| | | "Sig. (2-tailed)" | . | .000 |
| | | N | 32 | 32 |
| | Grey_Literature | "Correlation Coefficient" | 0.612** | 1.000 |
| | | "Sig. (2-tailed)" | .000 | . |
| | | N | 32 | 32 |
| **. "Correlation is significant at the 0.01 level (2-tailed)" | | | | |

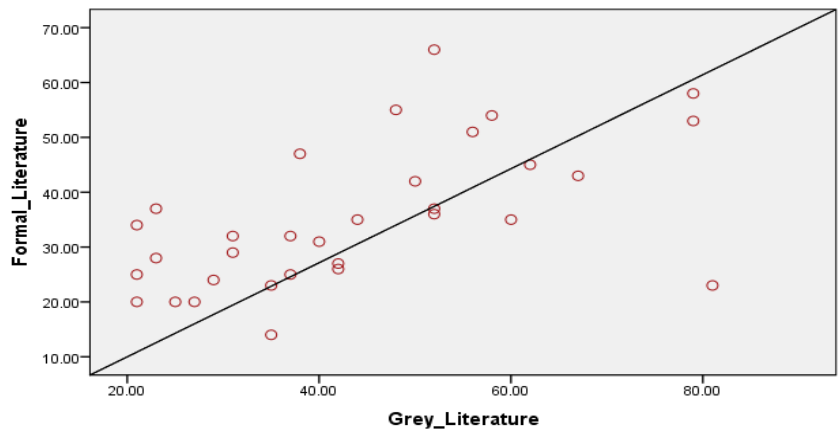

Figure 4: Scatter plot of success factor ranks of both data sets

Moreover, we applied the independent samples *t*-tests to assess the mean difference between both data sets (Tables 8 and 9). The *t*-test results (t = -2.200 and p = 0.062< 0.05) given in Table 8 illustrate that there are more similarities than differences between the ranks obtained from both data sets. Therefore, it shows a positive agreement between the findings from academic literature and the data collected from the grey literature.

Besides, we calculate the group statistics for both data sets to compare the mean difference between the data obtained for academic and grey literature. Group statistics results are shown in Table 9.

Table 8: Independent sample t-test

| | "Levene's Test for Equality of Variances" | "t-test for Equality of Means" |
|---|---|---|

|  |  | F | "Sig." | "T" | "df" | "Sig. (2-tailed)" | "Mean Difference" | "Std. Error Difference" | "95% Confidence Interval of the Difference" | |
|---|---|---|---|---|---|---|---|---|---|---|
|  |  |  |  |  |  |  |  |  | Lower | Upper |
| Rank | "Equal variances assumed" | 3.174 | .080 | -2.200 | 62 | 0.062 | -8.46875 | 3.84901 | -16.1628 | -.77469 |
|  | "Equal variances not assumed" |  |  | -2.200 | 57.057 | 0.061 | -8.46875 | 3.84901 | -16.1760 | -.76141 |

Table 9: Group statistics

|  | Group | N | Mean | Std. Deviation | Std. Error Mean |
|---|---|---|---|---|---|
| Rank | Academic data | 32 | 35.2188 | 12.93341 | 2.28633 |
|  | Grey data | 32 | 43.6875 | 17.51578 | 3.09638 |

### 4.4 Critical success factors

The critical factor presents the core areas where the organizational management needs to pay more attention to successfully manage the development activities [9, 33, 45] because they indicate the critical business areas. We find the criticality of each identified factor based on the criteria developed by [38] i.e.

- Factor declared as critical if its frequency of occurrence is ≥50%.

The same criteria are previously adopted by various researchers in other software engineering domains [31, 35, 45, 47]. Therefore, according to the frequency analysis (Table 3), the critically declared success factors of CGSD are:

*SF2 (continues development (around the clock), 55%), SF3 (time to market, 64%), SF5 (financial restructuring, 51%), SF8 (scalability, 50%), SF12 (trust building, 52%), SF15 (formalize relationship between overseas teams, 60), SF18 (information safety and security, 62%)* and *SF24 (reliable support structure, 54%).*

### 4.5 Mapping of identified success factor with PMBOK Knowledge Areas

Project Management Body of Knowledge (PMBOK), an industry de-facto project management guideline, has defined 10 knowledge areas that are recognized as good practices that can enhance successful management of a project [47]. Each knowledge area represents a set of core concepts and tasks that a project manager must consider to successfully manage project activities[48]. In this study, we mapped the identified success factors across the 10 PMBOK knowledge areas. We believe that this mapping will help a project manager to understand how the identified success factors are related to different aspects of managing CGSD projects.

Three researchers of the study were involved in the mapping activity in which we labeled and classified the related success factors into relevant PMBOK knowledge areas, as shown in Figure 5. A similar mapping process has been used by other studies (e.g.[49, 50]). We further conduct inter-rater reliability analysis to validate the mapping process. Moreover, three external experts were invited to take part in the inter reliability test. We calculate the 'non-parametric Kendalls coefficient of concordance (W)',[30] to calculate the inter-rater agreement between both the mapping teams (i.e. authors and external experts). The coefficient of concordance value (W=0.91 (p=0.003)) indicates agreement between the mappings of study researchers and external experts.

The mapping shows that majority of identified success factors are related to procurement knowledge area. Three success factors were mapped to risk, stakeholder, time, quality, and communication knowledge areas. Moreover, only one success factor namely 'integration with organizational IT infrastructure was mapped to the project integration management area.

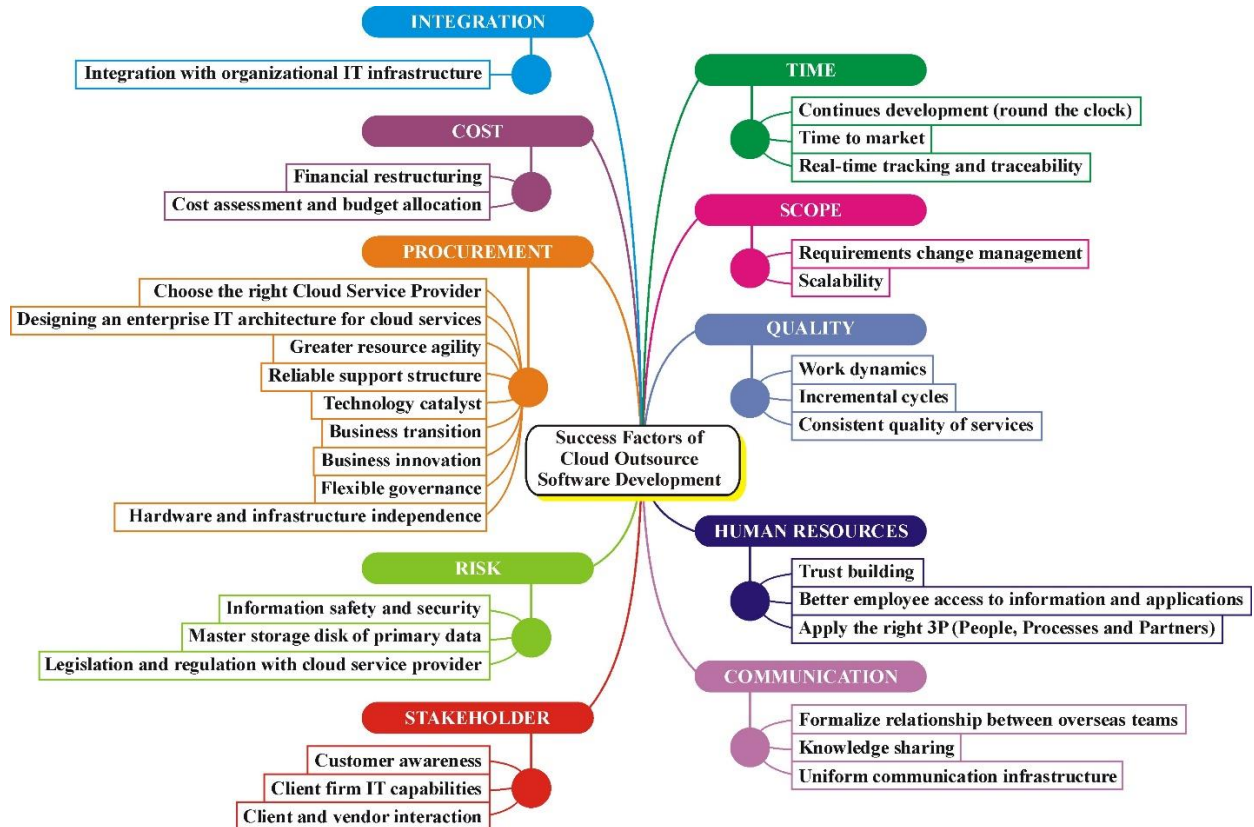

Figure 5: Theoretical framework of the investigated success factors

## 5. Summary and Discussion
### 5.1 Results Summary
The alternate aim of this research work is to develop a readiness model which assists the software organization to assess and improve the capabilities related to CGSD. This study focusses only a single component of the proposed readiness model i.e. success factors of CGSD.

**RQ1: What are the success factors for managing CGSD projects as identified in the multivocal literature review?**
Multivocal literature review study is conducted with the aim to explore the factors that could positively affect the CGSD practices. Using the step-by-step protocols of MLR, we finally selected 132 formal studies and 52 pieces of grey literature. The selected literature material was further considered for data extraction process and a list of 32 success factors was identified. The reported success factor highlights the key areas of the CGSD, which practitioners need to focus for successfully managing the development actives across the geographical boundaries.

**RQ2: Are there any differences between the success factors identified in the formal and grey literature?**

The multivocal literature review approach has been adopted, where both formal and grey literature data is collected to explore the success factors of CGSD. The formal literature studies published in academic research and grey literature are practitioner's experiences reports, blogs and case studies etc. The t-test and Spearman rank-order correlation test is applied to measure the statistical differences between the ranks of the reported factors for both formal and grey literature data sets. The t-test (t = -2.200 and p = 0.062< 0.05) and correlation ($r_S$ (32) =0.612) results discussed in section-4.3 reveal that there are no significant differences between the identified success factors in both formal and grey literature domain.

**RQ3: What success factors are related to CGSD vendor or client organizations?**

The identified factors are further analyzed with respect to the types of CGSD organization. Considering the frequency of identified success factors, we performed the chi-square test to check the significant differences between the success factors related to the client and vendor CGSD organizations. Based on the chi-square test results, we found significant differences between three success factors with respect to client-vendor classification i.e.SF2 (continues development (around the clock), p=0.008), SF17 (uniform communication infrastructure, 0.041) and SF24 (reliable support structure, 0.034).

Moreover, we noted that SF5 (financial restructuring, 52%, and 49%), SF6 (cost assessment and budget allocation, 35%, and 32%), SF11 (consistent quality of services, 34%, and 32%), SF14 (apply the right 3P (people, processes and partners), 46% and 43%), SF19 (master storage disk of primary data, 30% and 32%), SF29 (hardware and infrastructure independence, 30% and 30%), are the most common reported success factors in client and vendor organization.

**RQ4: What success factors are related to 10 knowledge areas of PMBOK?**

The identified success factors are further mapped in the 10 knowledge areas of project management. The aim of mapping process is to classify the factors related to their specific knowledge areas. The mapping results shows that majority of the success factors are scale in the procurement knowledge area.

Summary of the research questions results is provided in Table 10.

Table 10: Summary of the research questions

| Research questions | Findings |
| --- | --- |
| RQ1: What are the success factors for managing CGSD projects as identified in the multivocal literature review? | Total 32 factors are identified that could positively impact the global software development activities across the cloud platform. The list of the identified factors is provided in Table 4. |
| RQ2: Are there any differences between the success factors identified in the formal and grey literature? | Based on the frequency of occurrence in formal and grey literature, we found a positive correlation rs (32) = 0.612. In addition, the results of the t-test highlight that there are no significant differences in both data sets with respect to investigated success factors: (t = -2.200 and p = 0.062< 0.05). |
| RQ3: What success factors are related to CGSD vendor or client organizations? | The investigated success factors are categorized with respect to client and vendor organizations. The results demonstrate that there are significant differences between the client and vendor organizations for only three success factors, i.e., SF2 (Continuous development, p=0.008), SF17 (Uniform |

| | communication infrastructure, 0.041) and SF24 (Reliable support structure, 0.034). |
|---|---|
| RQ4: What success factors are related to 10 knowledge areas of PMBOK? | We classify the identified success factors into ten knowledge areas of PMBOK. The results shows that 'procurement' is declared as the most significant knowledge area of investigated success factors. |

## 5.2. Study Implications and Future work

The findings discussed in this paper makes contributions to the cloud based global software development literature. First, in this study, we have carried out a multivocal literature review to identify the success factors for managing CGSD projects. The success factors are identified from both formal published literature and industry grey literature. The study also analyzes the success factors with respect to both client and vendor CGSD organizations' perspectives. We believe that the identified success factors and their mapping to PMBOK 10 knowledge areas will act as a knowledge base for the CGSD research community. Moreover, the findings of the study will enhance the awareness of success factors associated with managing CGSD projects and will help researchers develop new strategies and frameworks to better manage the CGSD projects.

The study also makes contribution to practical implications in the domain of CGSD projects. For example, the study indicates that procurement is one of key knowledge area of managing CGSD projects. Project managers need to focus on selection of cloud service provider, tailoring enterprise IT architecture to support cloud service and develop flexible governance models to manage CGSD projects. Similarly, project managers need to focus on managing risk, communication, time, and human resources knowledge areas to better manage a CGSD project. Moreover, practitioners need to develop strategies to better manage changing requirements in a CGSD project. In a nutshell the study provides a deep overview of both academic and grey literature on CGSD project management, which has not been conducted before.

The ultimate future of this study is the development of a readiness model for CGSD (RMCGSD) organizations that will help the practitioners to measure and improve the cloud based global software development activities. The complete architecture of proposed model (RMCGSD) is provided in Figure 6 and it is based on the ideas of existing readiness models (like, SOPM[51], SOVRM[45], and CMMI[52]) and the factors that could impact the CGSD process. The RMCGSD consists of three core components, i.e., readiness level component, factors component (critical success factors (CSFs), critical barriers (CBs)) and assessment component. Figure 6 shows the association between the key components of RMCGSD. The readiness level component considered to assess the readiness level of a firm regarding the CGSD process, and the factors component consists of the CSFs and CBs that represent the key areas of CGSD process. The assessment component used to assess the specific readiness level of an organization and suggest the best practices to boost the CGSD capabilities of a firm.

Finding of this study will be used as one input to develop the readiness model, i.e., success factors of CGSD. Furthermore, we also plan to conduct a MLR study to identify challenges and best practices associated with CGSD projects. To develop the readiness levels of proposed RMCGSD, we will conduct SLR, MLR and questionnaire survey study to collect the barriers of CGSD. We also plan to conduct MLR and empirical study (interview and questionnaire survey) aiming to collect the effective best practices against each identified critical success factor and barrier. At final stage, once we develop the readiness levels of RMCGSD, we will conduct case study with different organizations to check the implacability of RMCGSD in real-world industry. The same model development process is used other software engineering domains [35, 45, 53].

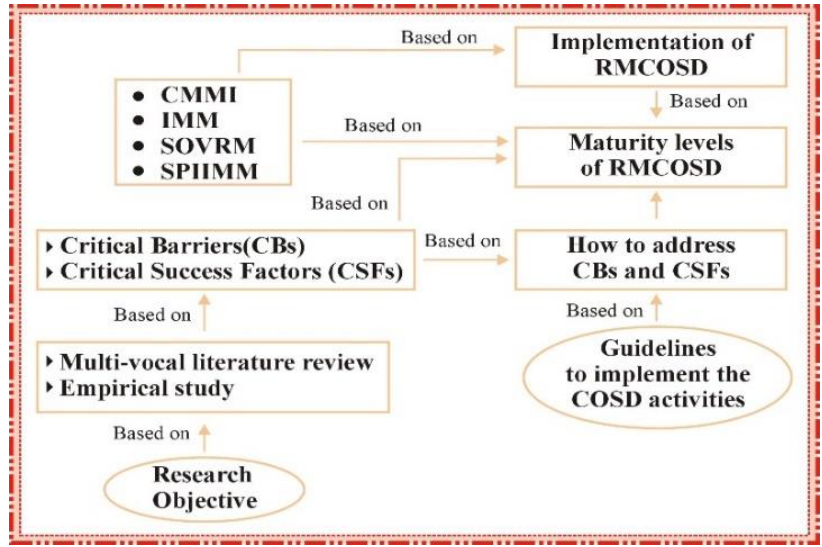

Figure 6: Structure of proposed model

## 5.3. Study Limitations
One potential threat toward the validity of this study findings is the incompleteness of a multivocal literature review. The findings of this paper are based on the literature extracted by using search keywords, limited digital databases, and search engines. To address this limitation, we use the alternatives of the keywords to develop a strong search string. Moreover, we used different digital databases and search engines to explore maximum literature related to the study objectives. Another possible limitation towards the validity of the study findings is the researcher's biases in identifying and mapping the success factors into PMBOK knowledge areas. To address this threat, we performed an inter-rater reliability test to examine and minimize the researcher's biases.

## 6. Conclusion
The objective of the study is to identify the success factors which explicitly affect the global software development activities using the cloud platform. We conducted an MLR study and identified 32 success factors exploring 184 primary studies. Moreover, the identified factors are further analyzed with respect to client and vendor CGSD organizations. The study shows that there are significant differences between the three success factors, i.e., '*continuous development (around the clock),*' '*uniform communication infrastructure*' and '*reliable support structure.* We further map the identified factors with client and vendor organizations considering their frequency of occurrence. We found that out of 32 factors, 20 success factors are more relevant with vendor organizations and 11 are with client organizations. The results shows that SF18 (Information safety and security) is equally important for both types of organizations. We further conducted a ranked based analysis based on the data collected form formal and grey literature, aiming to check which success factors is highlight reported in what category of data. The statistical results, correlation ($r_s$ (32)=0.612) *t*-test (t = -2.200 and p = 0.062< 0.05), there is no significant difference in frequency of occurrence of success factors in both types of data sets. Finally, the identified success factors are also scaled across the ten knowledge areas of the PMBOK framework. The mapping results show that procurement is the most significant knowledge area of the investigated success factors that need special focus of practitioners for successful CGSD projects.

**Appendixes**
**Appendix-A:** Selected, formal primary studies (https://tinyurl.com/y49oqw5t).
**Appendix-B:** Selected grey data sources (https://tinyurl.com/y6m2h297).
**Appendix-C:** Grey data collected through personal contact (https://tinyurl.com/yy3zt7ng).